\documentclass[conference]{IEEEtran}
\IEEEoverridecommandlockouts
\usepackage{cite}
\usepackage{amsmath,amssymb,amsfonts}
\usepackage{algorithmic}
\usepackage{graphicx}
\usepackage{textcomp}
\usepackage{xcolor}
\def\BibTeX{{\rm B\kern-.05em{\sc i\kern-.025em b}\kern-.08em
    T\kern-.1667em\lower.7ex\hbox{E}\kern-.125emX}}

\begin{document}

\title{
Cortex Inspired Learning to Recover Damaged Signal Modality using ReD-SOM Model \\
\thanks{Authors thanks 3IA Cote d'Azur fundings. Funding's code: ANR-19-P3IA-0002}
}

\author{\IEEEauthorblockN{Artem R. Muliukov, Laurent Rodriguez and Benoit Miramond}
\IEEEauthorblockA{\textit{Laboratoire d'Electronique, Antennes et Télécommunications (LEAT)}\\ \textit{Université Côte d'Azur}\\
Sophia Antipolis, France  \\
firstname.surname@univ-cotedazur.fr}
}

\maketitle
 
\begin{abstract}
Recent progress in the fields of AI and cognitive sciences opens up new challenges and problems that were previously inaccessible to study. One of such modern tasks is recovering lost data of one modality by using the data from another one. A similar effect (called the McGurk Effect) has been found in the functioning of the human brain. Observing this effect, one modality of information interferes with another, changing its perception. In this paper, we propose a way to reproduce such an effect and use it to reconstruct lost data modalities by combining Variational Auto-Encoders, Self-Organizing Maps, and Hebb connections in a unified ReD-SOM (Reentering Deep Self-organizing Map) model. We are inspired by human's capability to use different zones of the brain in different modalities, in case of having a lack of information in one of the modalities. This new approach not only improves the analysis of ambiguous data but also restores the intended signal. The results obtained on the multimodal dataset show an increase of quality of the signal reconstruction. The effect is remarkable both visually and quantitatively, specifically in presence of a significant degree of signal's distortion.
\end{abstract}



\section{Introduction}

\subsection{Biological inspiration}

It is no secret that human brain functioning is largely multimodal \cite{smith2005development}. It involves the study of the surrounding world by analysing multiple sensory representations of objects, often by quite independent brain zones. Deep information analysis occurs in the most advanced parts of our brain, particularly in the cortex. The cortex is divided into different zones responsible for processing different sensory information and solving different cognitive tasks \cite{cappe2009multisensory}. Therefore, despite their independence, the zones are widely cross-connected. Different zones of the cerebral cortex can influence each other's reasoning, as, for example, it was shown in the McGurk effect \cite{macdonald1978visual}. The most famous case is when duplicating a voice pronouncing [b], on a person's face pronouncing [g], the sound [d] is heard by observers. This phenomenon is often called the fusion effect \cite{tiippana2014mcgurk}. In this case, the perception of auditory information is influenced by additional visual information.

Although this effect can disturb the perception of a signal, it can also help to better understand a damaged signal, by complementing an unclear sample with information from another modality. Such an effect was of great interest to us, so we decided to reproduce it in an artificial neural network structure, applying it to the problem of damaged signal reconstruction.

\subsection{AI model concept}

In this work, we use a model inspired by the self-organizing mechanisms observed in the cerebral cortex, namely the Self-Organizing Map (SOM) model \cite{kohonen1990self}. Advancing earlier researches on how multiple SOMs may work together, we propose a new solution for learning and using them in practice.

We are interested in the complementary extraction of information from several modalities in the context of unsupervised learning. This work provides a solution that allows the neural network (NN) to correct signal data of one modality thanks to another one, based on previously learned multimodal intrinsic data structure.

We explore a new NN architecture that combines the ideas of Variational Auto-Encoder (VAE) \cite{kingma2013auto}, Deep SOM \cite{forest2021deep} and ReSOM (Reentry SOM) \cite{khacef2020brain, muliukov2022unified}. In this work, each modality is processed by its own AE combined with SOM. Also, each SOM neuron is connected to neurons in the other modalities by direct connections, trained on the basis of the neurons co-activations. In this way, we reproduce the principal rule of Hebbian theory \cite{hebb2005organization}. Thus, the co-activation of neurons (and so their relationship to each other) is preserved in their synaptic connections for further cross-modal communication and correction.

Our model allows to capture cross-modal data dependencies using mechanisms inspired by the functioning of biological systems. Despite a wide range of possible applications for the model, in this work we focus on its capabilities to handle various data alterations: such as confusing multimodal signals, noisy signals and data occlusions. The paper is organized as follows. In section II we give quite versatile review of previous works which inspired or directly influenced this work. In section III we describe the model and its differences to previously published architectures. Section IV presents our experiments and results, showing the model's capacities. In section V, we discuss the model's innovations, limits and possible improvements.

\section{Related works}

Working in a rapidly growing field we will try to cover a wide range of papers, focusing on the important ones for understanding this work and it's evolution.

\subsection{Self-organizing cortex-like learning models}

Pioneering works in cortex inspired NN learning was written by T. Kohonen \cite{kohonen1982self, kohonen1990self}. In his works, Kohonen proposed to model the cortex self-organization mechanism using a 2-dimensional grid of neighbouring artificial neurons. Later this approach was significantly distilled to a simplified "Kohonen network" or the Kohonen's Self-Organizing Map (KSOM) \cite{kohonen2007kohonen}. However, this model has several significant drawbacks, such as the inefficiency of its application to specific classification problems, the SOM boundaries limits, SOM weights "freezing" closer to the end of learning, etc. \cite{fritzke1995growing, rougier2011dynamic, khacef2020brain}. So numerous works have tried to upgrade the classical algorithm by proposing some exciting ideas.

The KSOM updating rule is time-dependent, and it becomes less and less sensitive to structural changes in the data to the latest learning epochs. To keep the SOM flexible at any moment of learning, the Dynamic SOM (DSOM) \cite{rougier2011dynamic} exchanges the time dependence with proximity between neuron vector and signal. Thus the plasticity of the model is inversely proportional to the good representation of the data structure. The Pruning Self-Organizing Maps for Cellular Hardware Architectures (PCSOM) \cite{upegui2018pruning} suggests cutting dubious in-network connections to organize the neuron groups in significant clusters. The Growing SOM (GSOM) model proposes to grow the SOM during its training until a more appropriate network size is reached \cite{fritzke1995growing}. Our model is inspired by all those and other works but it uses another version of the self-organizing rule discussed in the following subsections.

\subsection{Auto-Encoders for SOM model}
SOM manipulates simple vectorized features, which must be suitable for direct generalisation and interpretation. Hence, its functioning is highly dependent on an intelligent features extraction procedure \cite{khacef2020improving}. 

Staying in the unsupervised paradigm, some researchers have suggested jointly train the Auto-Encoder (AE) \cite{kramer1991nonlinear} and the SOM models to upgrade their common performance. Some works have proposed statistical methods incorporating the SOM loss into a global optimised AE loss \cite{pesteie2018deep, fortuin2018som}. A recent work suggests to process sequential data by using this approach \cite{huijben2022som}. Some other researchers suggest an integrated version of classical Kohonen's update formula as a loss function for the back-prop optimisation process \cite{forest2021deep}. We use a similar paradigm for defining the optimisation function, finding it more convenient for our model.

Several works suggest alternative ways of using the SOM to aggregate sub-parts of images within the SOM \cite{mohebi2014convolutional,aly2020deep,hankins2018somnet,braga2020deep}. But still, all of them stay unimodal and do not look at the problem from the angle of signals/images reconstruction perspective.

Nonetheless, some works mix supervised and unsupervised SOM learning to gather data of different nature. For example, in \cite{riese2019supervised} the authors tackle the reconstruction problem but in an unimodal and supervised way, compared to our work.

\subsection{Reconstruction models}

Numerous methods are proposed in the domain of signals reconstruction \cite{ahishakiye2021survey, montalt2021machine} : including problem-oriented analytical methods, signal filtering and decomposition methods, and data-driven and learning methods. The last ones are often gradient-descent centred, based on other popular architectures such as the GANs\cite{goodfellow2020generative}, the VAE \cite{kingma2013auto} or the U-net \cite{ronneberger2015u}. As an example, we can mention the work \cite{sadeghi2021mixture} solving the problem of signal reconstruction using multimodal data and VAE, but without usage of Hebb's connections and SOM.

Earlier developed models have already achieved outstanding results. Yet, in this work, we try to develop some new bio-inspired ideas. I.e., the usage of the SOM network and Hebb's connection rule for multimodal data reconstruction. The ideas we develop here are consistent with the previously proposed methods and could often be used jointly to improve and/or acquire new functionalities.

Several works have proposed SOM-based approaches to solve the signal reconstruction problem, mostly in surface reconstruction domain \cite{chuang2007application, kumar2004curve, yoon2008self, lamrini2011data}. However, they did not use the joint learning of SOM Feature Extractor (FE) leaning on the classical KSOM model paradigm.

\subsection{Multi-modal self-organizing models}

A more recent work that explores direct connection of multimodal self-organized neurons is the Reentry SOM (ReSOM) framework \cite{khacef2020brain, muliukov2022unified}. The ReSOM used raw data in the training stage, and it did not involve the joint training of a Feature Extractor (FE). Another work \cite{rathi2018stdp} uses a concept similar to the neural reentry, but adapted to the Spiking NN (SNN) domain. It was proposed as a preliminary work on simple data and it was also ignoring the stage of FE training. 

An alternative method to connect different SOMs is to add a Convergence Divergence Zone (CDZ) \cite{meyer2009convergence} that acts as a grid-mapper for clusters of neurons of all modalities in one place. The method is rather popular and various works have proposed its application, mainly in the domain of robot's orientation in space \cite{lallee2013multi, escobar2016self}. A CDZ paradigm-based architecture may be trained in a way proposed in this paper, but we do not try to tackle the question in this paper.

\section{Model description}

\subsection{Unimodal case}

Our model consists of several neural blocks, each of which should be discussed separately. We start with a unimodal functioning mode, and then extend it to the multimodal mode.

\subsubsection{VAE}

The basic building block of the model is a Variational Autoencoder \cite{kingma2013auto}. We are interested in both its ability to create a compressed representation and its variational ability to reconstruct a stable signal even for shifted encoded representation.

In our work, we do not focus on the enumeration of possible encoder models and stay on a quite simple one. We use four convolutional layers architecture both for the encoder and for the decoder parts. This is enough to demonstrate the model's functioning. Nevertheless, a more tricky encoder model can help to solve more complex problems and work with more sophisticated datasets.

\subsubsection{SOM}

The encoded vectors are used to create a logico-spatial map representing the variety of a train dataset structure. For this purpose, the standard SOM model uses a 2d spatial grid of neurons. Each neuron stores a vector equal in dimension to the encoded signal. Those neurons represent typical data objects trained on the statistics of previously received signals. A unique feature of this clustering method is that similar objects are grouped and located next to each other due to the self-organization (SO) process.

Precisely the SO process of the standard SOM model consists of the following steps:

1. For each new signal representation, we find the most similar neuron within the SOM, later called the Best Matching Unit (BMU);

2. Next, the vectors of the BMU and its neighbours are corrected in the direction of greater similarity to the received signal according to the formulas (now and later throughout the paper, sometimes we modify originally introduced notations, keeping formulas uniform.):

\begin{equation}
\label{eq:KSOM_apd}
v_i = v_i + \epsilon\exp(-\frac{||C_i - C_{BMU}||^2}{\alpha exp(-\frac{T}{\eta})}) (z - v_i)
\end{equation}
or 

\begin{equation}
\label{eq:DSOM_apd}
v_i = v_i + \epsilon||z - v_i||\exp(-\frac{||C_i - C_{BMU}||^2}{\eta^2||z - v_i||^2}) (z - v_i)
\end{equation}

for SOM and DSOM, respectively. Here $z$ is the encoded input vector, $T$ - is a time dependent temperature parameter, $C_i$ - is the coordinate of neuron $i$ in the SOM, $\alpha$ and $\eta$ - normalising coefficients, $\epsilon$ - optimisation step and $v$ are the SOM representing neurons vectors. The strength of the correction depends on the Gaussian distribution coefficient, with the physical centers in the BMU map's location.

After repeating a large number of iterations through all accessible data samples, we get a map similar to the one presented in Fig. \ref{fig:mnistsom} for the case of training on the MNIST dataset. To differ this type optimisation with the gradient descent, later we call this method as iterative optimisation.

\subsubsection{Dense SOM layer in VAE} \label{dense_som_layer}

The inconvenience of mixing iterative optimisation and gradient descent (GD) based approaches prompted us to move to an alternative optimisation method. 

We look at the SOM as an intermediate fully-connected layer of a NN, but optimised using a non-standard loss function. The earlier proposed model for joint use of VAE and SOM at a similar way \cite{forest2021deep}, uses the following loss formula: 

\begin{equation}
\label{eq:DeSOM_loss}
L'_{som} = \alpha\sum{}\exp(-\frac{||C_i - C_{BMU}||^2}{T^2}) ||z - v_i||^2, 
\end{equation}

where $\alpha$ is decreasing with temperature $T$ learning rate. When it is differentiated (necessary to carry out the GD step), we get an equation that largely reminds the standard SOM optimisation rule (\ref{eq:KSOM_apd}), presented above.

In our work, we propose a slight edition of the loss formula (\ref{eq:ReDSOM_SOMloss}), eliminating the last time-dependent parameter - temperature $T$ (analogically to the DSOM algorithm):

\begin{equation}
\label{eq:ReDSOM_SOMloss}
L_{som} = \sum{}\exp(-\frac{||C_i - C_{BMU}||^2}{\frac{||z - v_i||^2}{\eta}}) ||z - v_i||^2.
\end{equation}

In such a way, the loss formula loses its iterative optimisation nature and becomes simpler to optimise using classical GD method. Moreover, the model becomes more dynamic and capable of re-adapting neurons in case of a sudden change in the functioning of the feature extractor.

\begin{figure}
  \centering
  \includegraphics[width=0.48\textwidth]{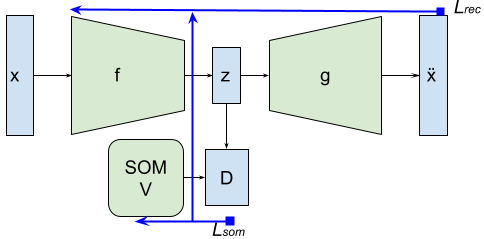}
  \caption{Deep SOM model \cite{forest2021deep}. Green
blocks are trainable objects, light blue - intermediate instances, blue lines are back-propagation paths. $x$ - input signal, $z$ - it's encoded version, $\ddot{x}$ - reconstructed signal. $f$ and $g$ - the encoder and decoder functions respectfully. $V$ - weights of SOM, $D$ - matrix of distances between $z$ and all vectors of $V$, needed to find the BMU and the $L_{som}$ function.}
  \label{fig:desomscheme}
\end{figure}

\subsubsection{Common unimodal loss}

The complete end-to-end unimodal architecture
is represented in Fig. \ref{fig:desomscheme}. It assumes an optimisation of a common loss function responsible for both: the quality of the signal reconstruction and the quality of the SOM approximation:

\begin{equation}
\label{eq:ReDSOM_unimloss}
L_{unimod} = L_{rec} + L_{som}
\end{equation}
here, $x$ stands for the original signal and $\ddot{x}$ the reconstructed signal.

\begin{equation}
\label{eq:ReDSOM_recloss}
L_{rec} =  ||x - \ddot{x}||.
\end{equation}

\subsection{Multimodal case}

\subsubsection{Hebbian connections}

The last important part of the model is the lateral connections, provided with a Hebbian learning law, which allows the training in the multimodal scenario. Similarly to this work \cite{muliukov2022unified}, we suggest to connect each pair of SOMs with that type of connections. Each pair of neurons from different SOMs has a weighted synaptic connection, which is trained by their statistical co-activation. In standard CNN domain terms, such connections can be called fully-connected layers.

\subsubsection{Common architecture}

\begin{figure}
  \centering
  \includegraphics[width=0.48\textwidth]{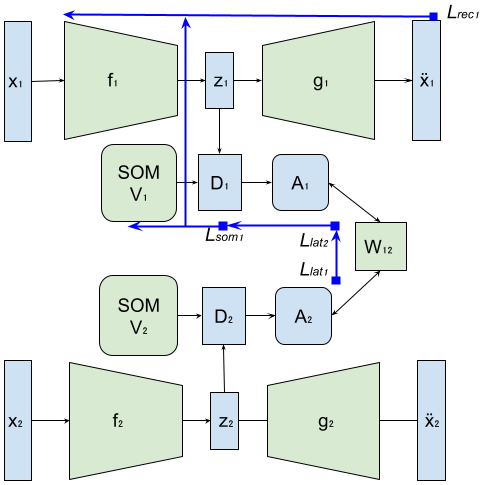}
  \caption{ReD-SOM model: data propagation scheme for 2 modalities. Green blocks are trainable objects, light blue - intermediate instances. Black arrows - data propagation, blue ones - later back-propagation. $x$ - input signal, $z$ - it's encoded version, $\ddot{x}$ - reconstructed signal. $f$ and $g$ - the encoder and decoder functions respectfully. $V$ - weights of SOM, $D$ - matrix of distances between $z$ and all vectors of $V$. $A$ - activation matrix and $W$ is Hebb's connections weights. With dark blue squares we indicate points of computing for different loss functions: $L_{rec}$, $L_{som}$ and $L_{lat}$.}
  \label{fig:redsom2mods}
\end{figure}

The general architecture of the model assumes the presence of k blocks, each composed of a couple: VAE and SOM. The SOM sub-blocks are connected among them by $\frac{k(k-1)}{2}$ Hebbian connections. To demonstrate the process of passing data through the model, we add a schematic presentation in Fig. \ref{fig:redsom2mods} for the case where the model processes 2 modalities data.

The model training is conducted by optimising a common loss function, which consists of three components ($L_{rec}$; $L_{som}$ and $L_{lat}$). The loss function is responsible for stabilizing the 3 key blocks (encoders/decoders, SOM and lateral connections).

\subsubsection{Multimodal aggregation}

The model takes a sum of three types losses among all accessible modalities $k_i$ and optimises it by GD algorithm as a single entity:

\begin{equation}
\label{eq:ReDSOM_ComLoss}
L = \alpha_{1}\sum_{k_i}L_{rec(k_i)} + \alpha_{2}\sum_{k_i}L_{som(k_i)} + \alpha_{3}\sum_{k_i, k_j}L_{lat(k_i, k_j)},
\end{equation}

where $\alpha_1$, $\alpha_2$ and $\alpha_3$ are normalising coefficients, defining the importance of each loss term for the learning. In a simplest case we can ignore the coefficients and take them equal to $1$, but we notice the importance of their range for the learning process. Hence the influence of their ratio deserves a separate research paper, and it was not explored in this very first work about the subject. Next we discuss separately the loss terms in next subsection.

\subsection{Loss functions}

\subsubsection{Reconstruction loss}

Reconstruction loss (\ref{eq:ReDSOM_recloss}) is responsible for correct reconstruction of each signal $x$ encoded into a vector $z$ ($z = f(z)$) and decoded into a vector $\ddot{x}$ ($\ddot{x} = g(z)$). A similar procedure is performed independently for each data modality.

\subsubsection{SOM loss}

The second loss function (\ref{eq:ReDSOM_SOMloss}) corresponds to the SOM construction. At the output of the encoder, we put a fully-connected layer $v$, which stores the grid of representing neurons, described earlier. 

Next, using formula:

\begin{equation}
\label{eq:ReDSOM_dist}
D_{i,j} = ||v_{i,j} - z|| 
\end{equation}

we find distances between the vector z and all $i,j$ vectors of layer $v$. After we find the BMU neuron/vector selecting the unit with the smallest distance:

\begin{equation}
\label{eq:ReDSOM_BMU}
BMU = \underset{i,j}{argmin}(D_{i,j}).
\end{equation}

Further, using the BMU, we can apply the optimised SOM loss function (\ref{eq:ReDSOM_SOMloss}). A similar process is conducted for each modality and can be done independently for different SOMs.

\subsubsection{Lateral loss}

The last loss is responsible for the optimisation of the Hebbian connections. First we define the activation map $A_i$ for each modality $i$:

\begin{equation}
\label{eq:ReDSOM_Act}
A^{i} = exp(\frac{-D'^2_{i}}{\sigma}),
\end{equation}

with $D'_{i}$ - normalised matrix of distances for modality $i$, found by (\ref{eq:ReDOM_DistNorm}): 

\begin{equation}
\label{eq:ReDOM_DistNorm}
D' = \frac{D - min(D)}{max(D) - min(D)}
\end{equation}

Here $\sigma$ is a hyper-parameter of the model. The idea of distance normalisation is borrowed from the work \cite{khacef2020brain}. It helps to make the activations distribution more uniform and homogeneous among different neurons.

Next we define the lateral activation coming from modality $k_i$ to $k_j$  (and the related lateral loss), produced by the Hebbian connections (weighted and represented by $W_{k_i,k_j}$). This lateral activation is defined in equation \ref{eq:ReDSOM_LatAct}.

\begin{equation}
\label{eq:RedSOM_ReDSOM_Latloss}
L_{lat(k_i, k_j)} = ||\tilde{A}_{k_i,k_j} - A_{k_j} ||
\end{equation}

\begin{equation}
\label{eq:ReDSOM_LatAct}
\tilde{A}_{k_i,k_j} = W_{k_i,k_j}, \times A_{k_i}
\end{equation}

Thus, the model trains the weights $W_{k_i,k_j}$ so that the lateral and afferent activations match to each other.

\section{Experiments and results}

\subsection{Data preparation and tests explanation}

\subsubsection{Multi-modal dataset creation}

To test the functionality in a multimodal scenario it was decided to combine some unimodal datasets, as in \cite{khacef2019written}. We use the MNIST dataset \cite{lecun2010mnist} and a part of the Google speech command dataset \cite{warden2018speech}. We specifically use a part which corresponds to the pronunciation of 10 numbers in English, later called Spoken MNIST or SMNIST. The audio dataset was transformed by Mel window transformation \cite{ittichaichareon2012speech} and represented in the form of 2-d images to simplify the VAE CNN application.  Dataset's instances were randomly sampled and merged so that the original object labels match. For further training, the corresponding labels were not taken into account. A part of the experiments was conducted with this synthetic 2-modalities dataset. Also, overall dataset was cut in 3 parts for train, test and the testing supervised oracle's learning (explained later in this section).

Then we experimented with the algorithm's functionality with a higher number of modalities. But due to the difficulty of finding and creating a more complex multimodal dataset, it was decided to artificially expand its multimodality with another visual unimodal dataset. For this, FMNIST \cite{xiao2017fashion} dataset was taken. This dataset contains grey colour images of clothes of 10 different types (classes), such as "t-shirt", "dress", etc. In the dataset natural labels are not presented as numbers, but this does not prevent us from matching numbers from 0 to 9 to their actual classes and supposing that those are the instances of new modalities. The association between digit representations and objects also brings some interesting features and questions about the model and its ability to associate objects, concepts and vocal labelling. 
These last experiments were conducted on this new three modalities dataset with suppression of actual data labels.

\subsubsection{Model's hyper-parameters}

\begin{figure}
  \centering
  \includegraphics[width=0.49\textwidth]{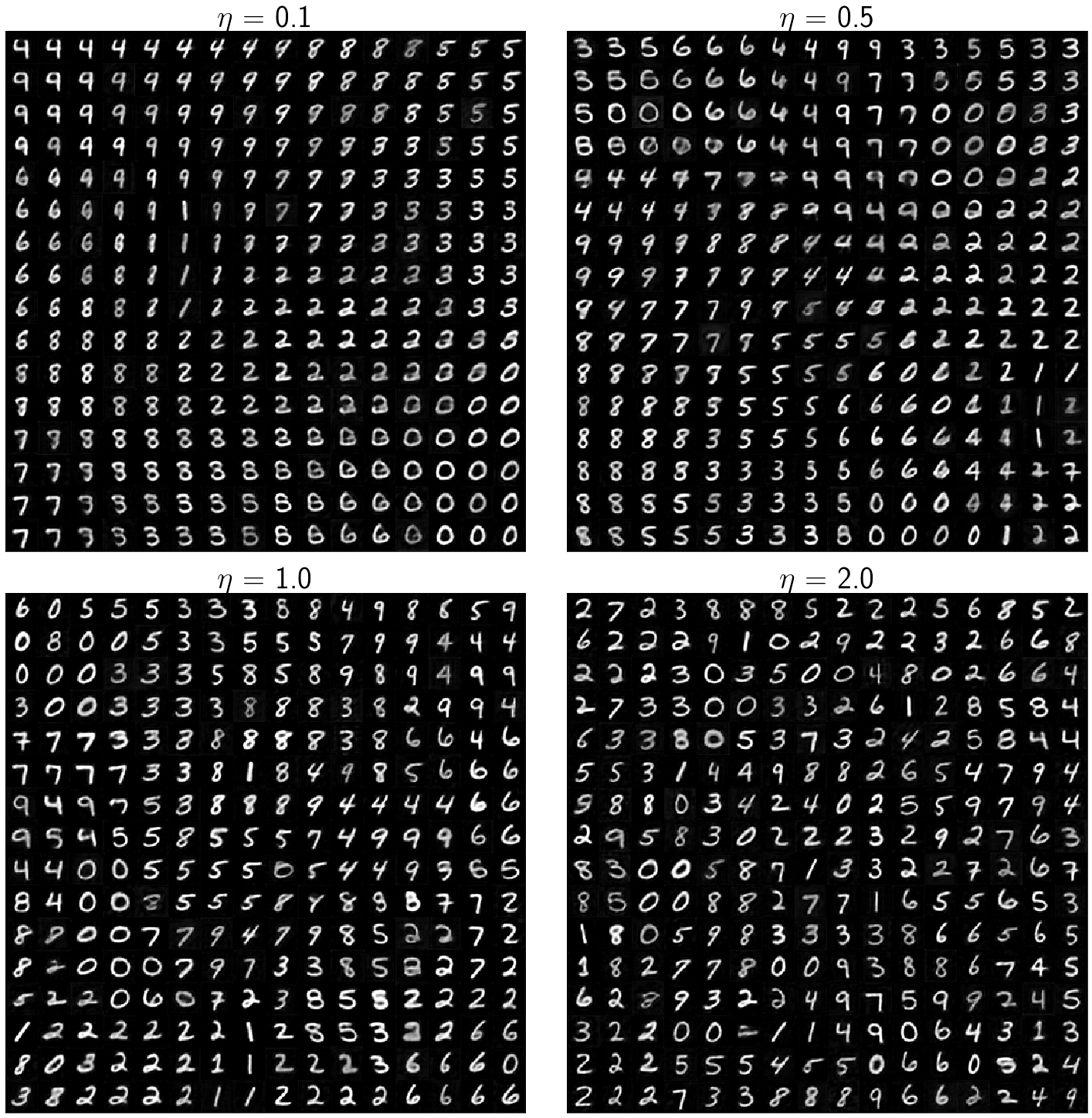}
  \caption{SOM map vectors reconstructed by decoder with different $\eta$ values. $\eta = 0.1$ close representing vectors but lower variability, $\eta = 2.0$ far vectors but high variability.}
  \label{fig:mnistsom}
\end{figure}

Even though our algorithm allows us to use any VAE model, we settled on a convolutional neural network (CNN) with four blocks for both the encoder and decoder (similarly to \cite{forest2021deep}). Each block of the neural network consists of a convolutional layer (or a deconvolutional one for the decoder), batch normalization, dropout, and leaky relu activation functions. Note that the search for the optimal NN architecture was not included in the scope of this work. We only discuss some of the VAE hyper-parameters (such as the number of convolutions for each layer of the size of $z$ vector). It can therefore be modified according to the complexity of the dataset and/or the available computational resources.

One important parameter of the model $\eta$ (from (\ref{eq:ReDSOM_SOMloss})) defines the level of proximity for SOM representations, reducing the neurons variety, making them closer to each other (Fig. \ref{fig:mnistsom}). For the reconstruction application we suggest setting an intermediate value (for ex. $\eta = 1$), keeping in mind that more challenging datasets may ask to have higher $\eta$ values, sacrificing the SOM objects proximity for the sake of higher representative neurons variability. 

Another parameter directly related to the data variability is the SOM space size $v$.  Keeping similarity with previously published works, we settled on a size of $16 \times 16$, which is large enough to interpret a significant number of classes but compact enough to visually analyse the built space. However, it can be changed depending on dataset complexity. 

A last parameter to discuss is the $\sigma$ from equation \ref{eq:ReDSOM_Act}. It defines how strong BMU neurons are activated compared to all other neurons. With  $\sigma = 0.3$ we have an intermediate activation level, noticing a bunch of the most strongly activated neurons.

\subsubsection{Experiments description}
\label{exp_desc}

{\bf A distinct feature of our model} is its ability to reconstruct signals, taking into account multimodal relationships between data presentations. So a damaged part of a signal may be filled up with an alternative presentation, selected by other modalities. Therefore in the presence of a large amount of distortion, the recovered signal can differ significantly from the original one. Nevertheless it recovered taking into account the multimodal data structure. Thus, the use of standard metrics to evaluate denoising quality (such as peak signal-to-noise ratio) is not a relevant evaluation method in this context.

We test our model on a dataset whose structure consists in the presence of common classes of object's multimodal presentation. Hence it is possible to evaluate the model's performance through analysis of preservation of the original class. So we have conducted both qualitative and quantitative study of this property.

{\bf First}, qualitatively, we show the model's capacities to handle confusing signals:  simple ones in unimodal case and more complex ones, up to completely ambiguous signals, in multimodal case. Next we show the model's ability to reconstruct noisy images in the presence of different types of noise, such as the Gaussian noise and the Salt\&Paper noise. Lastly, we show how the model behave on data occlusions of different types.

{\bf Second} we show a quantitative study of how the model performs in the presence of one of the selected noises, precisely the Salt\&Paper one. To conduct a statistically reliable study on a sufficient amount of test objects, we have trained in a supervised manner a strong NN classifier ResNet50 \cite{he2016deep} on an independent part of our dataset (of 14000 examples). Next we have measured how efficient the reconstruction is for recognition of its initial class by our ResNet50 oracle. The test is conducted on another 3500 couples of image and audio signals.

\subsection{Results}

\subsubsection{Unimodal case}

\begin{figure}
  \centering
  \includegraphics[width=0.48\textwidth]{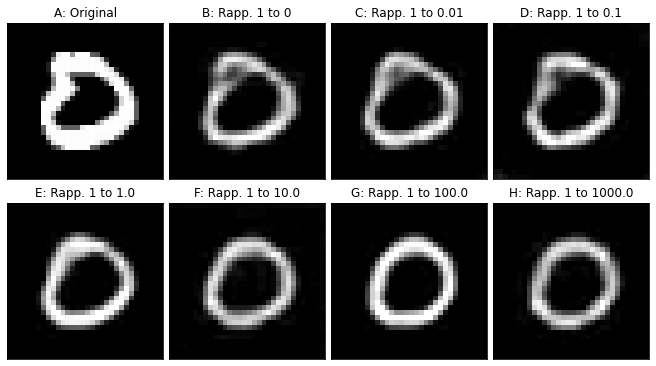}
  \caption{Standardisation of a MNIST image using unimodal signal in the ReD-SOM model. A - original image, B - reconstruction using only VAE model, C - H: Reconstructions of mixed $z$ + $V_{BMU}$ vectors from VAE+SOM model, with given ratio.}
  \label{fig:som_corr}
\end{figure}

First, let us consider unimodal architecture. Mixing values of the encoded $z$ vector and its corresponding BMU vector, we change the original signal encoding towards a more average version that corresponds to a simple signal correction method.  In this case, an object belonging to a well-defined class but on which an important artefact appears can be corrected. The new version of the object will be a more typical one, smoothed according to the past statistics. Such a transformation changing the form of a "0" image is illustrated in Fig. \ref{fig:som_corr}.

The method offers a way to choose the most reliable representations from the available SOM space, taking into account the model's experience.

\subsubsection{Multimodal case}

\begin{figure}
  \centering
  \includegraphics[width=0.48\textwidth]{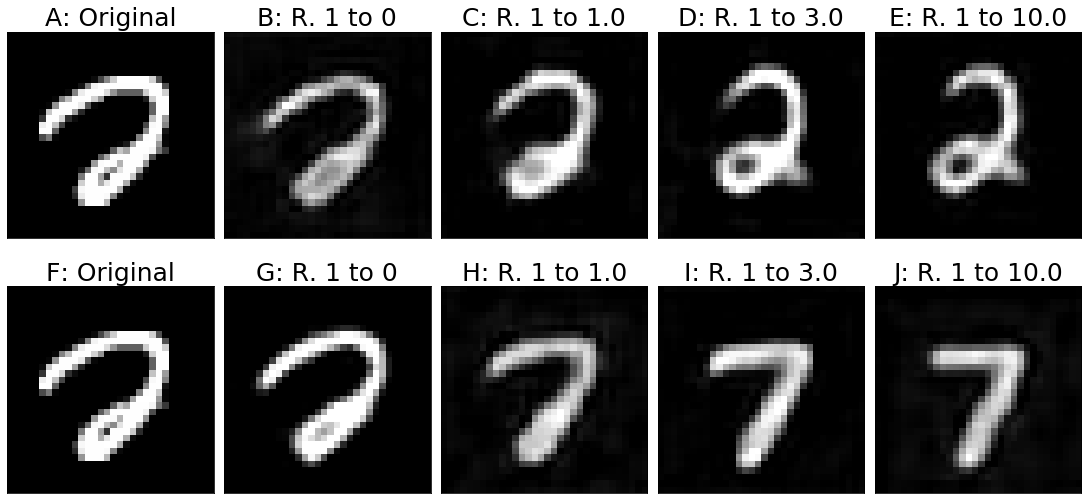}
  \caption{Drifting of image in different direction using sound signal thanks to multimodal association. Illustration with digit [two]([seven]) in upper line (lower line). A(F) - original image, B(G) - reconstructed version using only VAE, C - F (H-J): Reconstructions of mixed $z$ + $V_{lat-BMU}$ vectors, with given ratio (R).}
  \label{fig:resom_corr}
\end{figure}

The following method has similar characteristics to the McGurk effect \cite{tiippana2014mcgurk}, where one modality changes the perception of another one. In our model, we use the information of one modality (in Fig. \ref{fig:resom_corr}, the audio one) to correct the encoded $z$ vector in the direction of the information received in another data modality (the visual modality in Fig. \ref{fig:resom_corr}). 

To do so, we count lateral activations for all modalities $k_i$ and all neurons $j$, using equation \ref{eq:ReDSOM_LatAct}, and we find the most activated neuron according to \ref{eq:RedSOM_newAct}.  Next, we mix the current $z$ representation with the strongest lateral activated neuron-vector (\ref{eq:ReDSOM_newZ}).
  
\begin{equation}
\label{eq:RedSOM_newAct}
BMU_{lat} = \underset{k_i, j}{argmax}(A_{k_i}^j)
\end{equation}

\begin{equation}
\label{eq:ReDSOM_newZ}
z_{new} = \frac{z + rV_{BMU_{lat}}}{1 + r}
\end{equation}

Here $r$ is the lateral importance coefficient, defining how strong the lateral modality will affect the initial signal. Next, this new $z_{new}$ representation may be used for future analysis or for the signal reconstruction. As shown in Fig. \ref{fig:resom_corr}, an indefinite object (ground truth label - "two") is shifted to the direction of "2" or "7", depending on the given sound representation of the object, [two] or [seven].

\subsection{Noise suppression and Data reconstruction}

\begin{figure}
  \centering
  \includegraphics[width=0.48\textwidth]{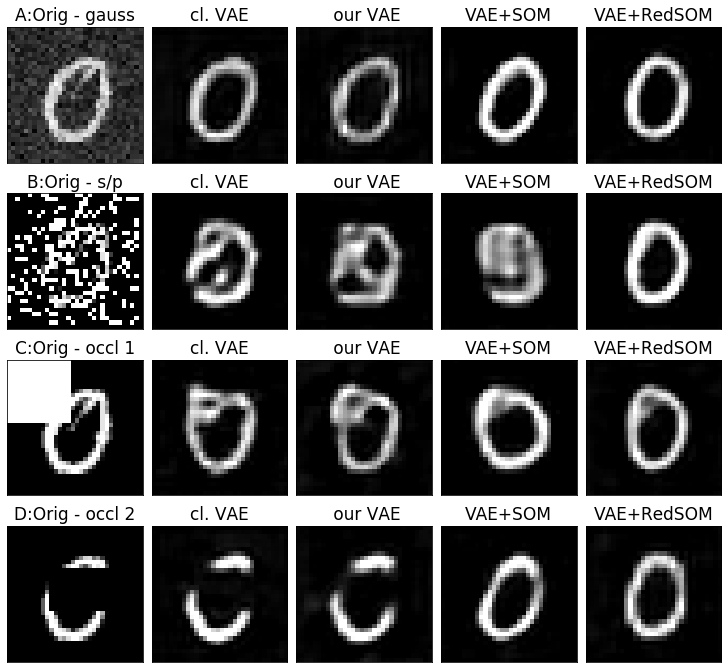}
  \caption{Correction of noisy images with different distortions: A - Gaussian noise, B - Salt and pepper, C -  partially overexposed image, D - partial lost signal. 1st col. - input signal, 2nd - classical VAE, 3rd - our VAE (trained within ReD-SOM model), 4th - VAE + it's SOM BMU with 1 to 1 ratio (unimodal method), 5th - VAE + lateral SOM activation with 1 to 1 ratio (using not corrupted signal from audio modality).}
  \label{fig:resom_noise}
\end{figure}

Next demonstration (in Fig. \ref{fig:resom_noise}) shows how the model can handle different types of signal distortions. The distortions might be of different nature, so we test some of them: Gaussian noise, Salt \& Pepper Noise, White and Black data occlusions. The model uses the SOM and the lateral SOM activations to find the best representing neuron and correct the $z$ encoding using formula (\ref{eq:ReDSOM_newZ}). 

We have compared different reconstruction methods: classical VAE; VAE trained in ReD-SOM model, but tested as a normal VAE; ReD-SOM unimodal and multimodal reconstructions. So as one can see on Fig. \ref{fig:resom_noise}, normally the reconstructions are visually closer to the original class "0" for each next method.

\subsection{Impact of the number of modalities}

\begin{figure}
  \centering
  \includegraphics[width=0.48\textwidth]{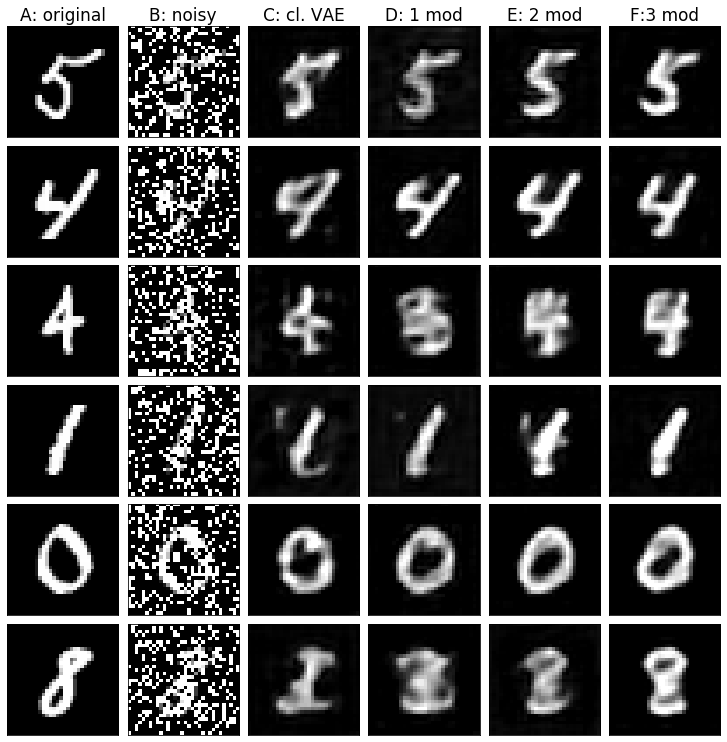}
  \caption{Examples of reconstruction using classical VAE and RED-SOM with different numbers of modalities. A - original images. B - noisy images, C - classical VAE reconstructed images, D - RED-SOM VAE reconstructed images, E - 2 modalities (MNIST and Spoken MNIST) RED-SOM reconstructed images, F - 3 mod. RED-SOM reconstructed images. }
  \label{fig:redsom_3mods_ex}
\end{figure}

The last study aims to examine the influence of the number of modalities on the quality of the correction. First we show the changes of reconstruction quality for different numbers of modalities (Fig. \ref{fig:redsom_3mods_ex}). Generally, the quality is improving with a number of modalities. But to confirm this fact we had to conduct a quantitative study.

For this purpose we have conducted two experiments for different types of signal: on MNIST dataset for visual signals and of SMNIST for audio signals. We have varied the amount of S\&P noise from 0\% to 90\% in the initial signal. We have applied all earlier described reconstruction methods, with some variants for multimodal datasets. For both experiments we tested three configurations: the MNIST dataset coupled with SMNIST, with FMNIST and with both of them. The SMNIST dataset respectively was coupled with MNIST, with FMNIST and also with both of them. 

Because a PSNR like metric is not relevant for this model (as discussed in section \ref{exp_desc}), we have used an oracle to control if the damaged input data is of the same class as the reconstructed one. The class prediction was accomplished by a deep ResNet50 \cite{he2016deep} oracle trained on a labelled unimodal dataset. We compared the predicted labels with the ground truth ones.

\begin{table}[t]
\centering
\caption{\% of errors in ResNet50 prediction for different reconstruction modes. Noise level corresponds to ratio of damaged neurons}
\begin{tabular}{|l|c|c|c|c|c|}
\hline
Noise level & 0 & 0.1 & 0.2 & 0.3 & 0.5\\
\hline
\hline
\multicolumn{6}{|c|}{MNIST} \\
\hline
Noisy data & 0.81 & 16.73 & 48.39 & 72.98 & 87.90\\
VAE & 1.41 & 3.83 & 5.85 & 9.48 & 29.03\\
REDSOM VAE & 1.61 & 4.03 & 7.26 & 12.30 & 34.88\\
REDSOM Stand. & 1.21 & 3.83 & 4.84 & 9.88 & 34.68\\
REDSOM MNIST+SMNIST & {\bf0.60} & 1.81 & 3.02 & 8.27 & 25.00\\
REDSOM MNIST+FMNIST & 1.01 & 2.22 & 3.43 & 5.65 & 21.98\\
REDSOM 3 mod & {\bf0.60} & {\bf1.21} & {\bf2.42} & {\bf5.24} & {\bf17.14}\\
\hline
\multicolumn{6}{|c|}{Spoken MNIST} \\
\hline
Noisy data & {\bf9.27} & 39.72 & 63.71 & 77.42 & 84.48\\
VAE & 26.41 & 36.69 & 46.98 & 56.85 & 74.60\\
REDSOM VAE & 23.39 & 35.89 & 48.39 & 54.84 & 73.39\\
REDSOM Stand. & 24.60 & 33.67 & 41.13 & 46.98 & 72.18\\
REDSOM MNIST+MNIST & 23.99 & 27.22 & 32.26 & 35.48 & 45.77\\
REDSOM MNIST+FMNIST & 26.01 & 32.86 & 38.71 & 38.10 & 50.60\\
REDSOM 3 mod & 20.77 & {\bf21.37} & {\bf25.20} & {\bf30.44} & {\bf32.86}\\
\hline
\end{tabular}
\label{tab:ReD-SOM_multimod}
\end{table}

As one can see in the Table \ref{tab:ReD-SOM_multimod}, unimodal data recovery using our method does not significantly affect the accuracy of data recognition. But the recognition of multimodally reconstructed data is statistically better than the ones reconstructed by the unimodal VAE. The effect is noticeable already when using two modalities and becomes even clearer when using three. So we can state that an increase in the number of modalities positively affects the reconstruction's quality. This is so for both experiments: for the visual reconstructed modality (MNIST) and for the audio modality (SMNIST).

\begin{figure}
  \centering
  \includegraphics[width=0.49\textwidth]{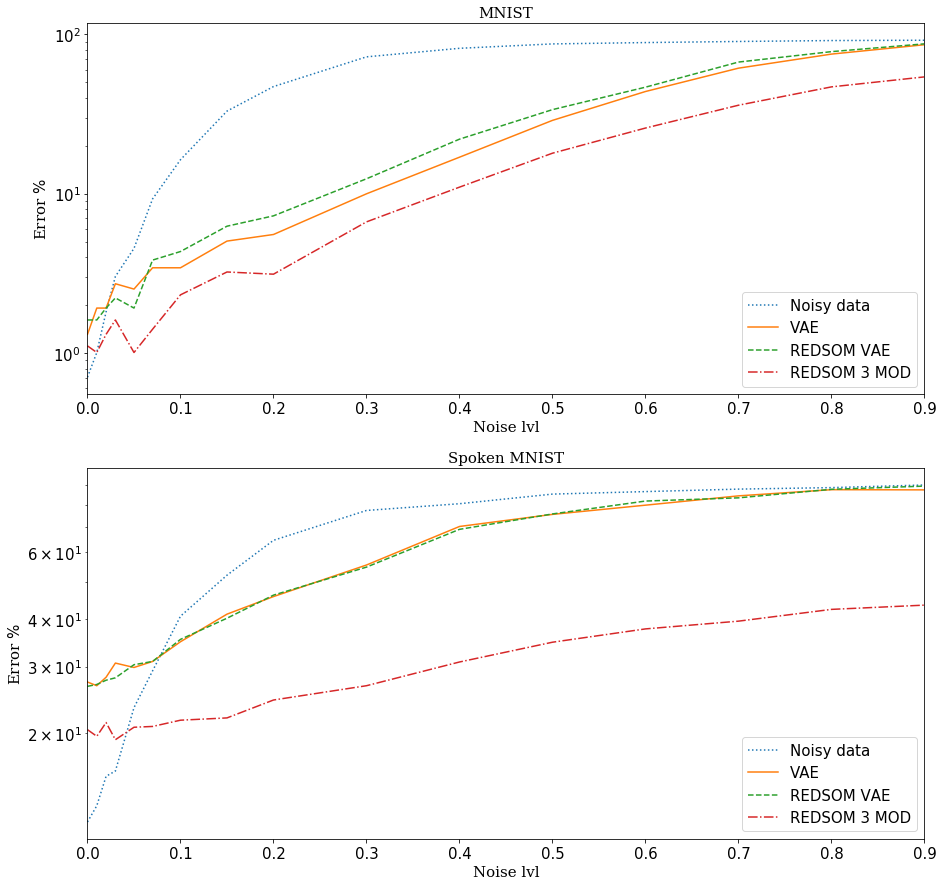}
  \caption{Prediction error rate in log. scale (for supervised ResNet50 trained on MNIST and SMNIST, respectively) of reconstructed signals depending on S\&P noise levels. Reconstruction methods are: VAE, in RED-SOM trained VAE and 3 modality RED-SOM VAE. }
  \label{fig:rec_3mods_graph}
\end{figure}

Similar behaviour can be observed for any amount of noise from 5\% to 90\% (line "REDSOM 3 MOD" in Fig. \ref{fig:rec_3mods_graph}). The exceptions are the starting points (less than 1\% of noise for MNIST and 5\% of noise for SMNIST), where the amount of noise is so small that it is better not to pass data through the AE bottleneck. Then it is better to use the raw data (line "Noisy data" in Fig. \ref{fig:rec_3mods_graph}). Probably the use of a more complex AE model should solve this problem.

\subsection{Computing time analysis}

All tests were performed on Intel(R) Core(TM) i9-9880H CPU and Nvidia Quadro T2000 GPU. The model has been coded in Python using \textit{pytorch} \cite{NEURIPS2019Paszke} framework. We conducted the joint loss optimisation defined in (\ref{eq:ReDSOM_ComLoss}). We have used an advanced version of the gradient descent algorithm, namely the Adam optimiser \cite{kingma2014adam}. We conducted 1000 learning epochs with full passage among 52500 unique multimodal representations of the training set. 

Full training procedure for the unimodal MNIST (Speech Commands reps.) model took 9.47 hours (9.83 hours resp.). Training of the most complete 3-modalities model took 29.23 hours. Thus, the addition of modalities proportionality multiplied the average training time. Note that the 3-modalities model's training time is comparable to the training time of the 3 independent unimodal networks.

In Table \ref{tab:comp_time} one can find consolidated data about the computing times. In 2 modalities cases the numbers 2(a) and 2(b) refer to different combinations of modalities, respectively a) MNIST+SMNIST for the first case, and b) MNIST+FNIST and SMNIST+FNIST for the 2nd one.

Reconstruction procedure for 1000 MNIST (SMNIST resp.) samples takes 3.52 $\pm$ 0.55 seconds (3.54 $\pm$ 0.52 seconds resp.) with one modality; 5.32 $\pm$ 0.97 seconds (4.99 $\pm$ 0.40 seconds) with two modalities; 6.81 $\pm$ 1.09 seconds (6.59 $\pm$ 0.70 seconds) with three modalities. Thus, addition of 2 modalities to the reconstruction process increases the computational time less than 2 times.

\begin{table}[t]
\centering
\caption{ Computing time for networks train and reconstruction procedure depending on number of modalities. }

\begin{tabular}{|l|c|c|c|c|}
\hline
mod. num. & 1 & 2a & 2b & 3 \\
\hline
\multicolumn{5}{|c|}{Training time in hours (for 1000 epochs)} \\
\hline
MNIST & 9.47 & 18.30 & 21.21 & 29.23 \\
SMNIST & 9.83 & x & 19.93 & 29.23 \\
\hline
\multicolumn{5}{|c|}{Reconstruction time in seconds (for 1000 samples)} \\
\hline
MNIST & 3.52$\pm$0.55 & 5.32$\pm$0.97 & 5.26$\pm$0.84 &  6.81$\pm$1.09\\
SMNIST & 3.54$\pm$0.52 & 4.99$\pm$0.40 & 5.14$\pm$1.01 & 6.59$\pm$0.70\\
\hline
\end{tabular}
\label{tab:comp_time}
\end{table}

\section{Discussion and Conclusion}

\subsection{Discussion}

The importance of a particular loss term for the correct signal reconstruction is not discussed in detail in our work, but it is quite an interesting question. By changing the hyper-parameters ratio $\alpha$ in the optimised function (\ref{eq:ReDSOM_ComLoss}), we can tune the model to be more sensitive to the preservation of the original signal, or rather try to preserve information about the multimodal cross-activation of neurons, up to ignoring the possibility of subsequent signal recovery.

Due to this property, the present model prioritizes preservation of the multimodal relationship between views over exact reconstruction of the signal.  As a result, the method cannot be considered as a means of precise signal reconstruction, but rather as a technique for preserving the contextual meaning of the signal during restoration.

Such a model can be useful in various applications, where most precise signal reconstruction is not in the focus. For example, when the problem of accurate signal reconstruction cannot be solved at all. The multimodal information can provide valuable information for tasks such as object recognition or scene understanding. For instance, this method can provide a worthy reconstruction system for decision-making and control tasks, for which any representation of the complete surrounding image may be useful.

\subsection{Conclusion}

We propose a new biologically-inspired idea in the domain of data correction. The proposed neural model enables a joint usage of Convolutional AE (CAE) and Multimodal Hebbian connections for better signal reconstruction. The model supposes statistical similarity of new data with earlier exposed ones and bases the reconstruction procedure on previously learned multimodal structure.

To resume innovations proposed in this work: 

\begin{itemize}
    \item we have introduced a new method of design for the SOM, simplifying its integration in classical NN models;
    \item we have demonstrated the possibility of joint train of SOM and CAE;
    \item we have analysed the reconstruction efficiency for different types and levels of distortions;
    \item we have shown that increasing the number of modalities
    can significantly improve the quality of reconstruction.
\end{itemize}

In conclusion, the current work has demonstrated a proof of concept for our idea. However, to further improve its performance, it is recommended to explore more advanced AE architectures with recurrent structures and improved reconstruction capabilities for complex objects. Further experiments using natural multimodal datasets, such as video-flows collected by multiple sensors, are necessary to validate the proposed approach.

\bibliographystyle{IEEEtran}

\bibliography{ref}

\end{document}